\documentclass[12pt]{article}
\usepackage{graphicx}

\newcommand\pubnumber{}
\newcommand\pubdate{\today}
\newcommand{\ttw}{$t\bar{t}W$}

\def\institute{Instituto de Física Corpuscular (IFIC)\\
Universitat de València -- CSIC, Valencia, SPAIN}
\def\support{\footnote{This project was funded by LCF/BQ/PI19/11690014}}

\def\Title#1{\begin{center} {\Large #1 } \end{center}}
\def\Author#1{\begin{center}{ \sc #1} \end{center}}
\def\Address#1{\begin{center}{ \it #1} \end{center}}

\newcommand\pubblock{\rightline{\begin{tabular}{l} \pubnumber\\
         \pubdate  \end{tabular}}}
\newenvironment{Abstract}{\begin{quotation}  }{\end{quotation}}
\newenvironment{Presented}{\begin{quotation} \begin{center} 
             PRESENTED AT\end{center}\bigskip 
      \begin{center}\begin{large}}{\end{large}\end{center} \end{quotation}}

\def\beq{\begin{equation}}
\def\eeq#1{\label{#1}\end{equation}}
\def\eeqn{\end{equation}}

\def\beqa{\begin{eqnarray}}
\def\eeqa#1{\label{#1}\end{eqnarray}}
\def\eeqan{\end{eqnarray}}

\let\bar=\overbar

\def\Dslash{\not{\hbox{\kern-4pt $D$}}}
\def\dslash{\not{\hbox{\kern-2pt $\del$}}}

\def\msb{{\bar{\ssstyle M \kern -1pt S}}}

\begin{document}
\begin{titlepage}
\pubblock

\vfill
\Title{\ttw\ Production: a very complex process}
\vfill
\Author{ Marcos Miralles López,\support \\ on behalf of the ATLAS Collaboration}
\Address{\institute}
\vfill
\begin{Abstract}
These Monte Carlo studies describe the impact of higher order effects in both QCD and EW \ttw\ production. Both next-to-leading inclusive and multileg setups are studied for \ttw\ QCD production.
\end{Abstract}
\vfill
\begin{Presented}
$13^\mathrm{th}$ International Workshop on Top Quark Physics\\
Durham, UK (videoconference), 14--18 September, 2020
\end{Presented}
\vfill
\end{titlepage}
\def\thefootnote{\fnsymbol{footnote}}
\setcounter{footnote}{0}

\section{Introduction}

The \ttw\ process is very interesting from the phenomenological point of view~\cite{Maltoni:2014zpa}. It is for instance a main background for some beyond the Standard Model (SM) searches and other rare top SM processes such as $t\bar{t}H$ and $t\bar{t}t\bar{t}$. Moreover, \ttw\ production rates have been measured at the LHC by CMS and ATLAS as inclusive cross--sections~\cite{Sirunyan:2017uzs,Aaboud:2019njj} and those measurements yield larger values than the SM predictions from the CERN Yellow Report 4~\cite{deFlorian:2016spz}. This motivates the in--depth study of this process.


For these Monte Carlo (MC) studies, the event selection is as follows: the $t\bar{t}$ pair is decayed semileptonically and the associated $W$ boson is decayed leptonically, being both leptons of the same charge. In addition, the following particle level jet cuts $p_T(j)>25$~GeV and $|\eta|<2.5$ are applied. Forwards jets are defined in the $2.5<|\eta|<4.5$ region. 

\section{Disentanglement of Higher Order Effects}

\begin{figure}[!htb]
      \centering
      \includegraphics[width=0.75\textwidth]{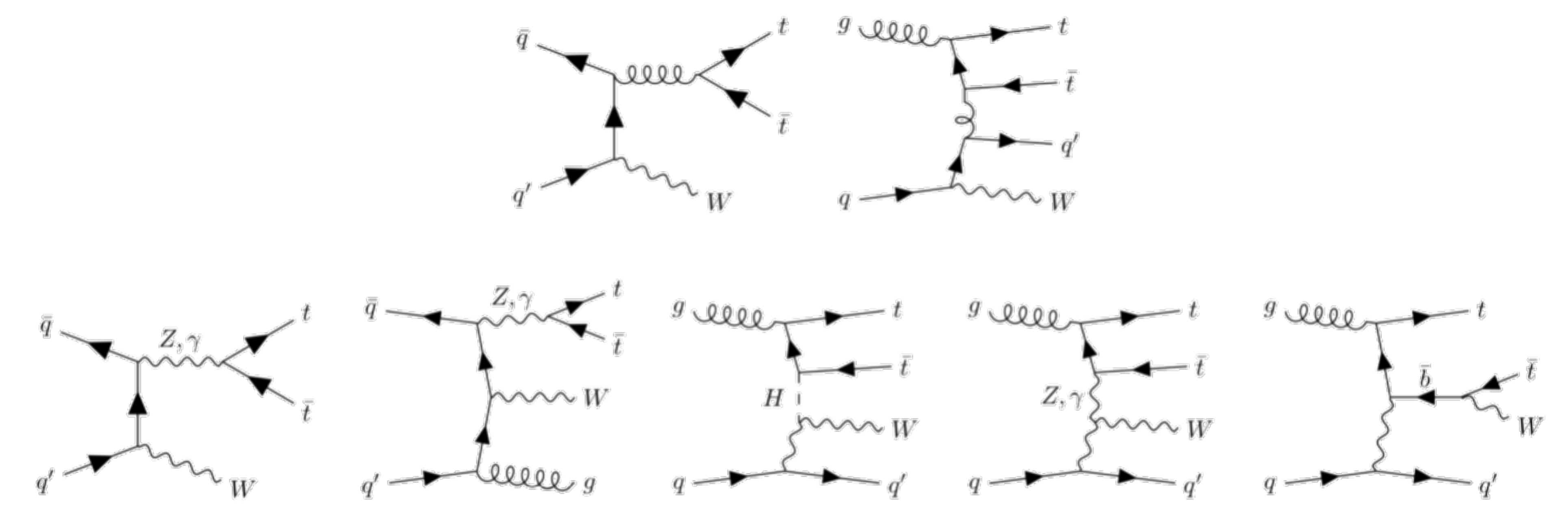}
      \caption{Examples of \ttw\ Feynman diagrams relevant for these studies: LO QCD $(\mathcal{O}(\alpha_S^2\alpha))$ and NLO QCD $(\mathcal{O}(\alpha_S^3\alpha))$ production in the top part, and the ``tree--level EW" contributions $(\mathcal{O}(\alpha^3+\alpha_S\alpha^3))$ in the bottom.}
      \label{fig:fig1}
\end{figure}

Higher order effects (in the quantum chromodynamic (QCD) strong coupling constant $\alpha_S$ and the electro--weak (EW) coupling constant $\alpha$) are very important for \ttw\ production and can significantly modify leading order cross--sections. Figure~\ref{fig:fig1} shows the Born level diagrams due to these higher order corrections that enter the MC simulations.

The \textsc{MG5\_aMC@NLO}~\cite{Alwall:2014hca} generator is used interfaced with the \textsc{Pythia8}~\cite{Sjostrand:2014zea} parton shower (PS) for both multileg and inclusive setups. The following items are explored: \textit{scale variations} of the renormalisation and factorisation scales $(\mu_R~\mathrm{and}~\mu_F)$ in the matrix elements (ME) (for inclusive setups), where up to three different functional forms are used; \textit{multileg setups} (with the FxFx~\cite{Frederix:2012ps} algorithm), using NLO--accurate matrix elements for up to one additional jet and LO--accurate matrix elements for up to two additional jets (\ttw\ $+0,1j$NLO $+2j$LO); \textit{parameter variations} that impact the FxFx matching algorithm.     

\section{QCD Production}

\begin{figure}[!htb]
      \centering
      \begin{tabular}{cc}
            \includegraphics[width=0.47\textwidth]{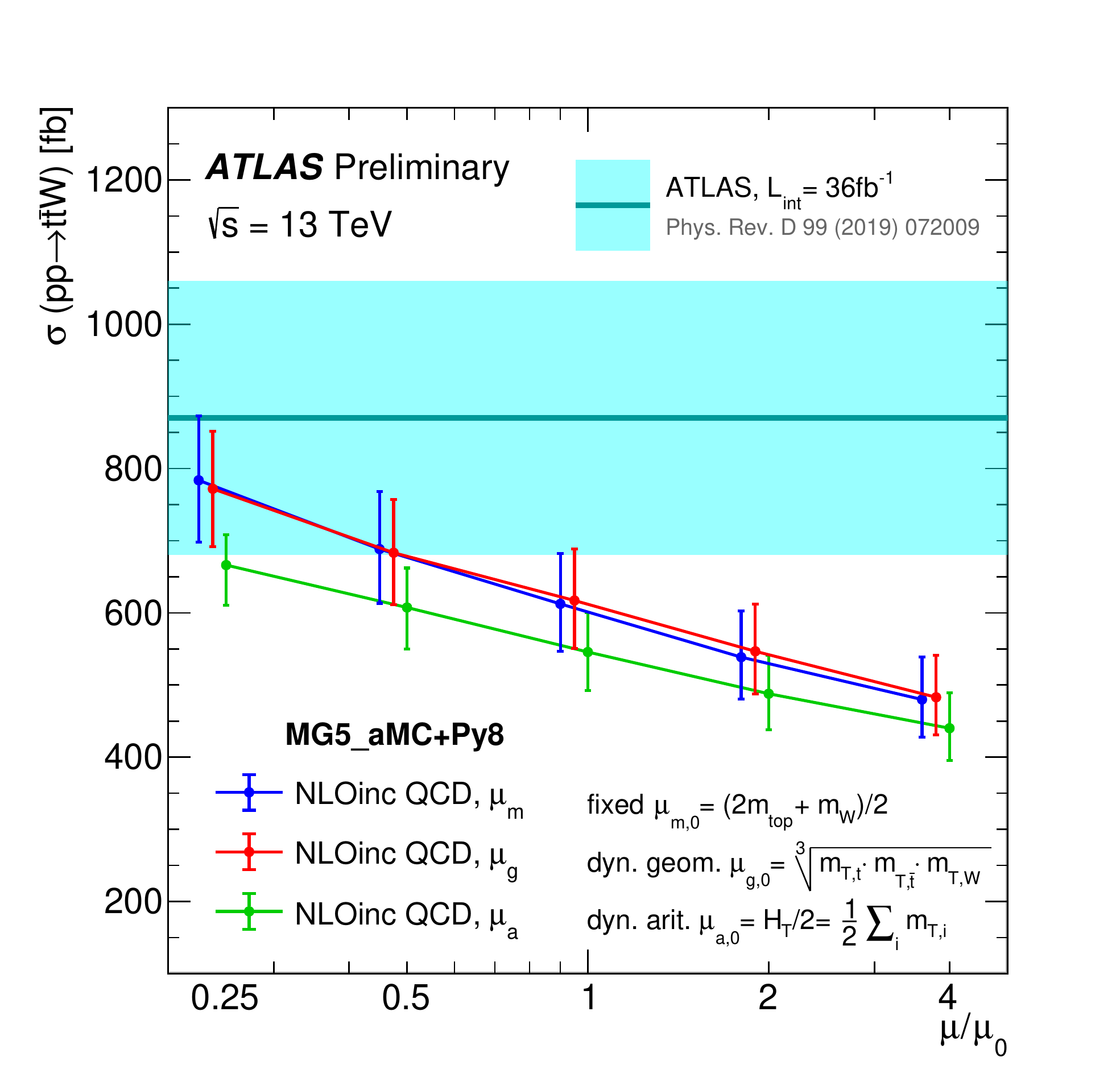} & 
            \includegraphics[width=0.44\textwidth]{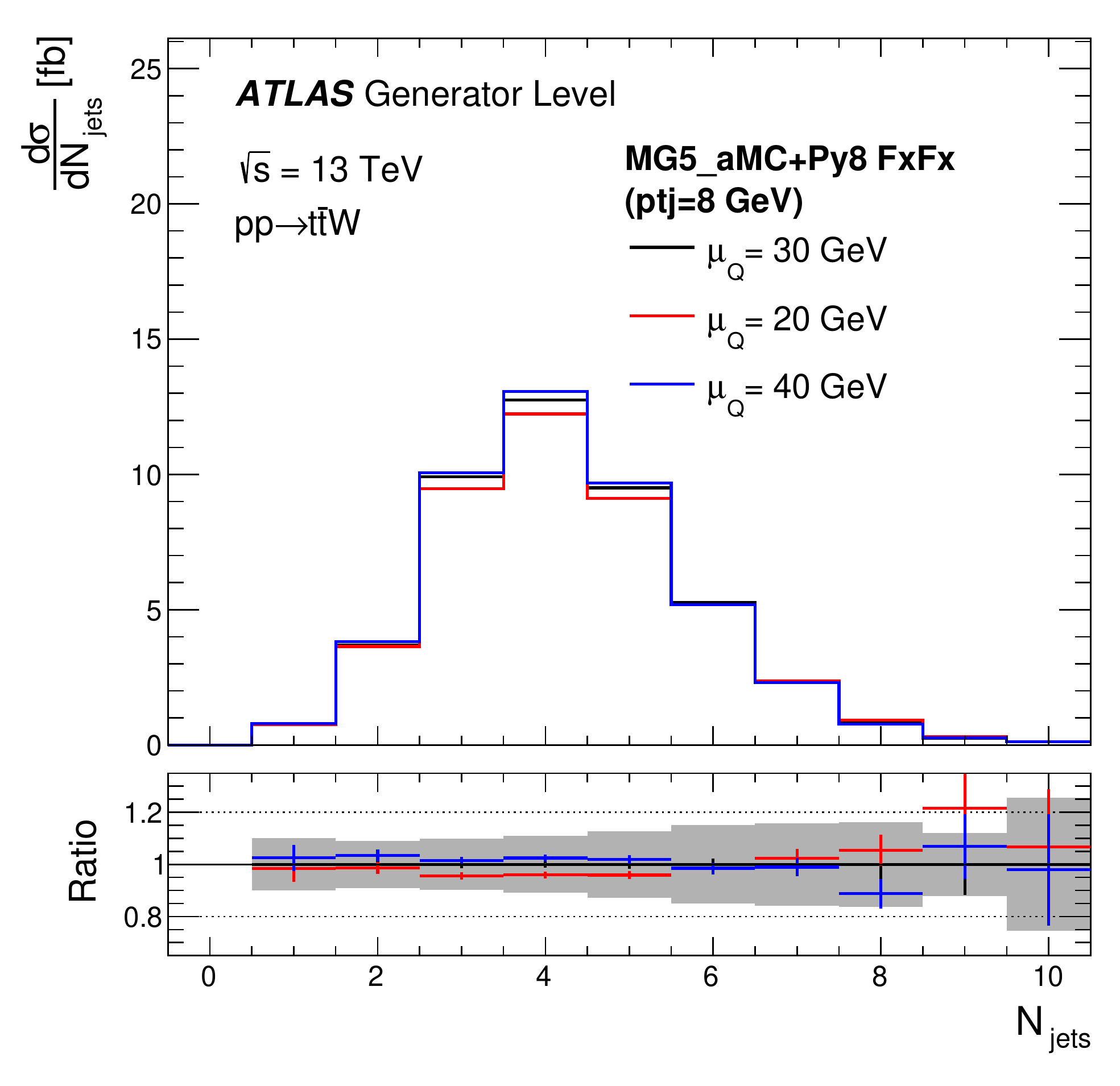} \\
            \footnotesize{(a)} & \footnotesize{(b)} \\ 
            \includegraphics[width=0.44\textwidth]{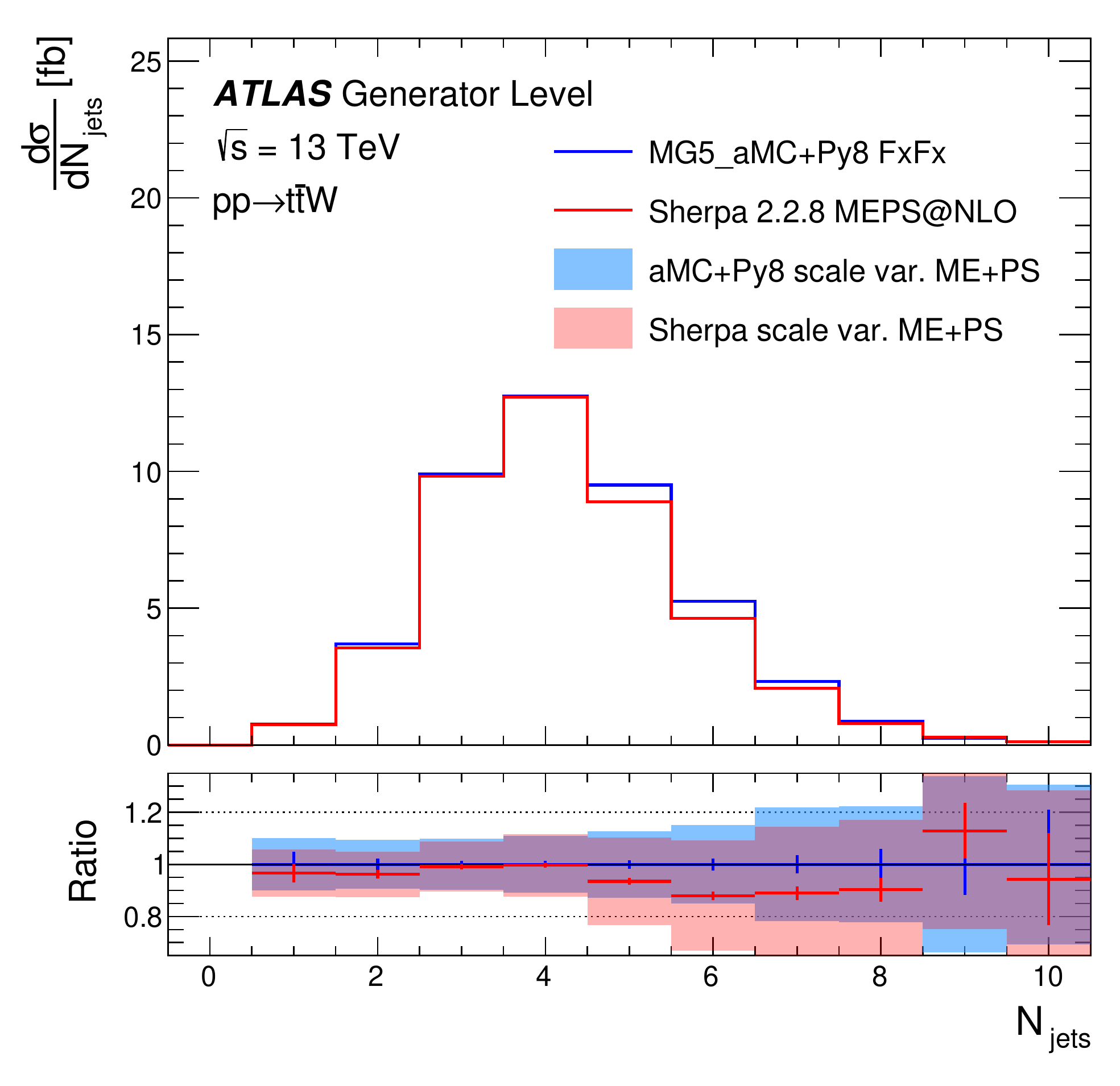} &
            \includegraphics[width=0.44\textwidth]{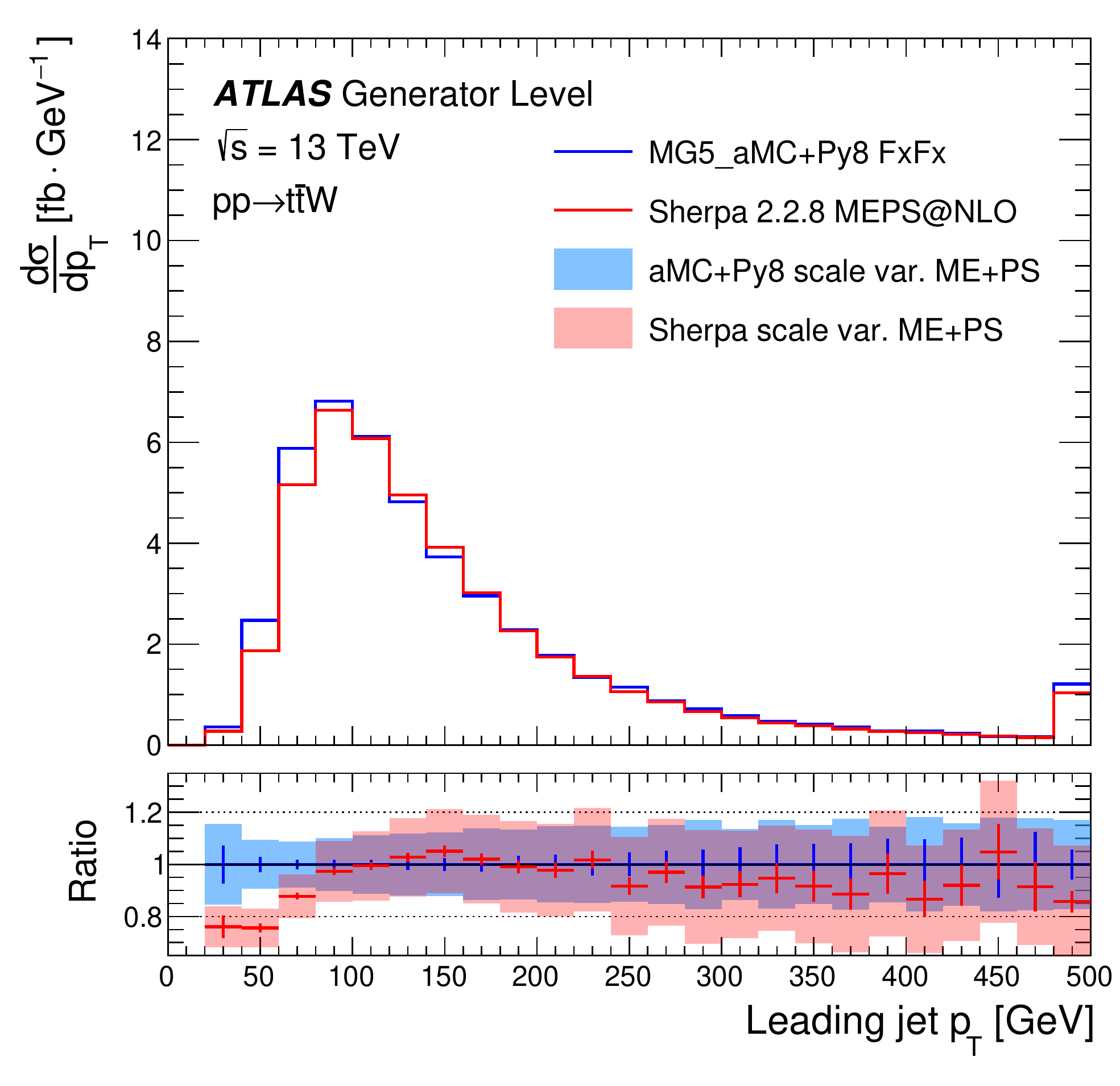} \\
            \footnotesize{(c)} & \footnotesize{(d)} \\
      \end{tabular}
      \caption{(a): Cross--section dependence with three functional forms of the $\mu_R~\mathrm{and}~\mu_F$ scales. (b): \textsc{MG5\_aMC@NLO} FxFx samples parameter variations. (c) and (d): Comparisons between the \textsc{MG5\_aMC@NLO} and \textsc{Sherpa2.2.8} MC generators. The vertical error lines show the 7--point scale variations for (a), while for the rest they indicate the MC statistical uncertainty and the shaded bands represent these scale variations. Figures from Ref.~\cite{ATLAS:2020esn}.}
      \label{fig:fig2}
\end{figure}

The QCD corrections have been studied with both NLO inclusive and multileg merged setups. Figure~\ref{fig:fig2} shows the studies performed at NLO QCD accuracy for the three points mentioned in the previous section. The following conclusions may be extracted:
\begin{itemize}
      \item (a): there is a $10\%$ increase in the cross--section between the (green) default dynamical scale used in MG5\_aMC@NLO and the (blue) fixed scale used in the CERN YR4. For all functional forms, there is a big dependence of the cross--section with the chosen value of the scale with $\sigma(\mu_{i,0}/4)/\sigma(\mu_{i,0})\sim1.4$.
      \item (b): the nominal multileg (FxFx) sample has a merging scale $\mu_Q=30$~GeV and a $p_T^{min}(j)=8$~GeV. No significant shape effects and a cross--section difference of about $2\%$ is observed when changing the merging scale. This configuration yields a cross--section of $\sigma_{t\bar{t}W}^{FxFx}=614.2_{-13\%}^{+12\%}$~fb.
      \item (c) and (d): there is good agreement between both \textsc{MG5\_aMC@NLO} and \textsc{Sherpa2.2.8}~\cite{Bothmann:2019yzt,Gutschow:2018tuk} multileg setups inside the uncertainty bands. These show correlated scale variations in the ME and PS.
\end{itemize}

\section{``tree--level" EW Production}

\begin{figure}[!htb]
      \centering
      \begin{tabular}{cc}
            \includegraphics[width=0.44\textwidth]{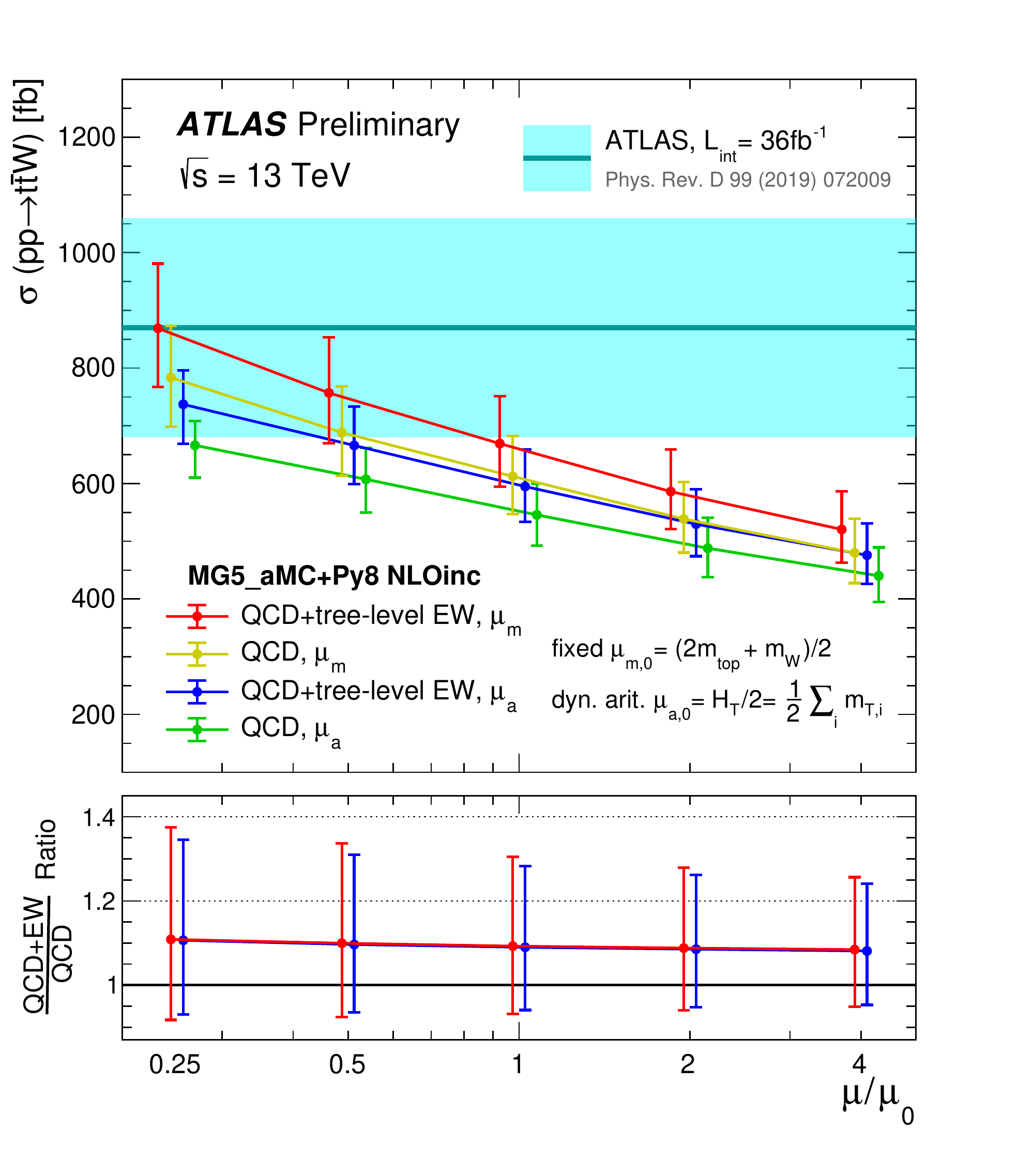} & 
            \includegraphics[width=0.44\textwidth]{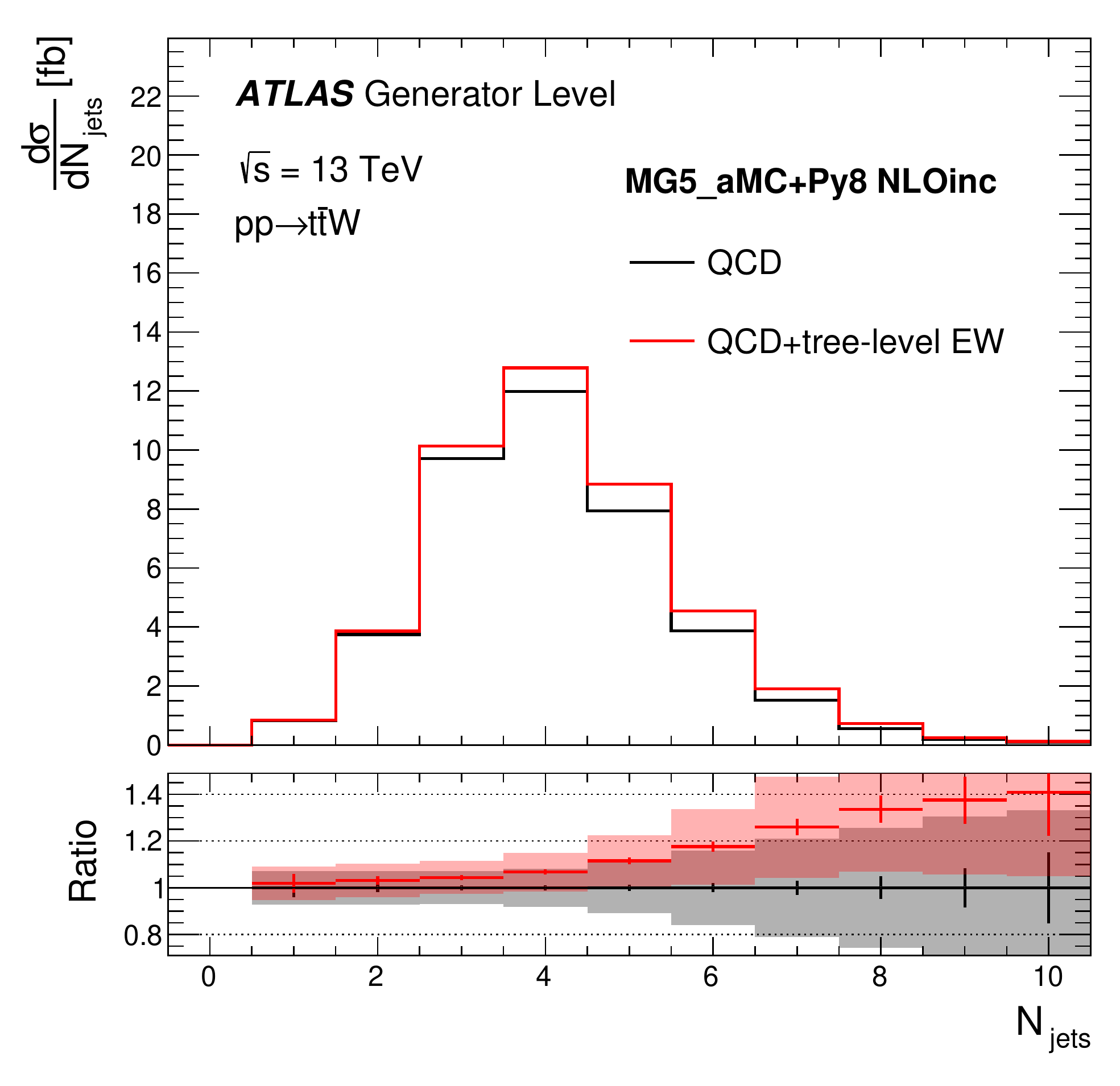} \\
            \footnotesize{(a)} & \footnotesize{(b)} \\ 
            \includegraphics[width=0.44\textwidth]{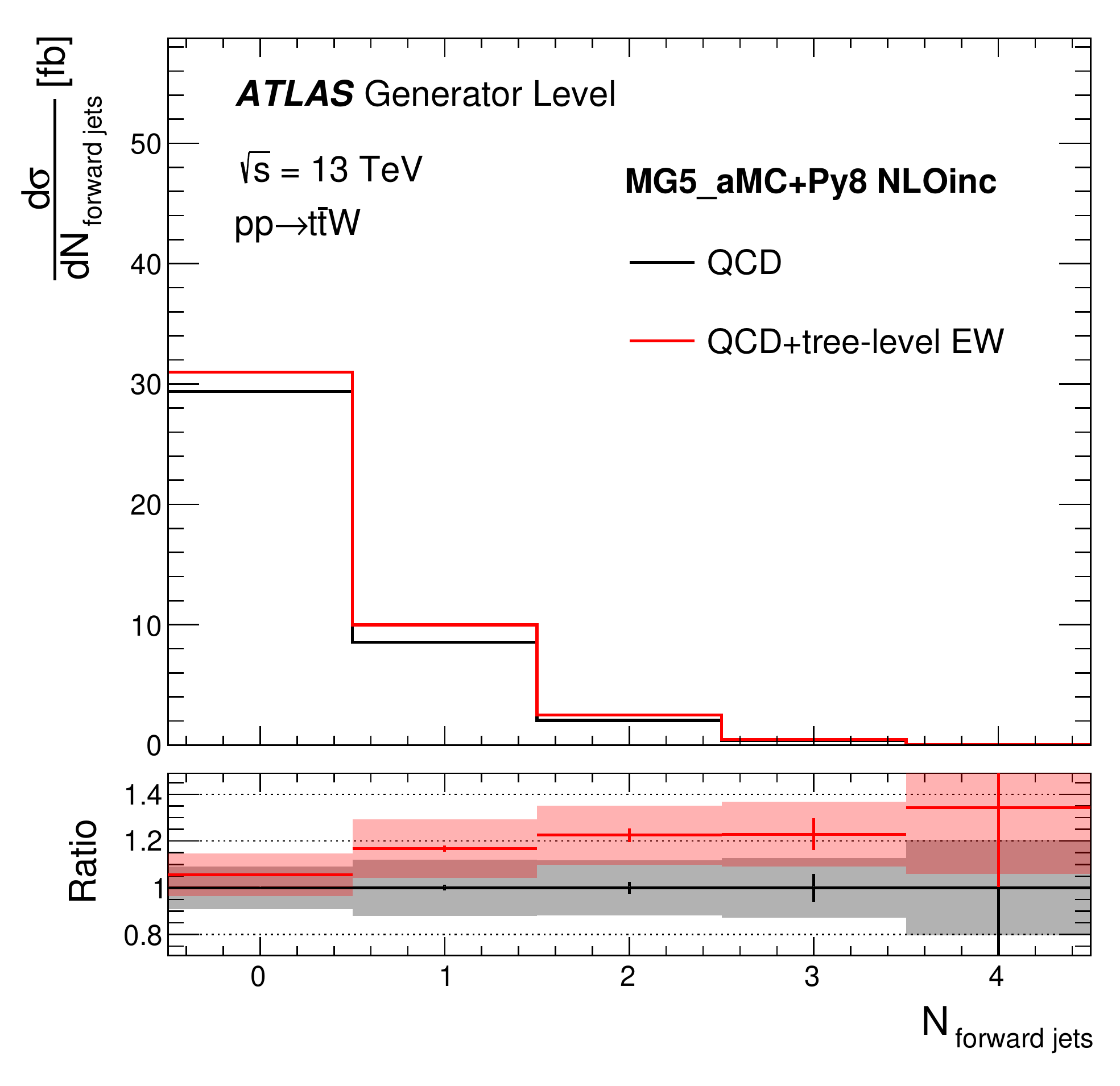} &
            \includegraphics[width=0.44\textwidth]{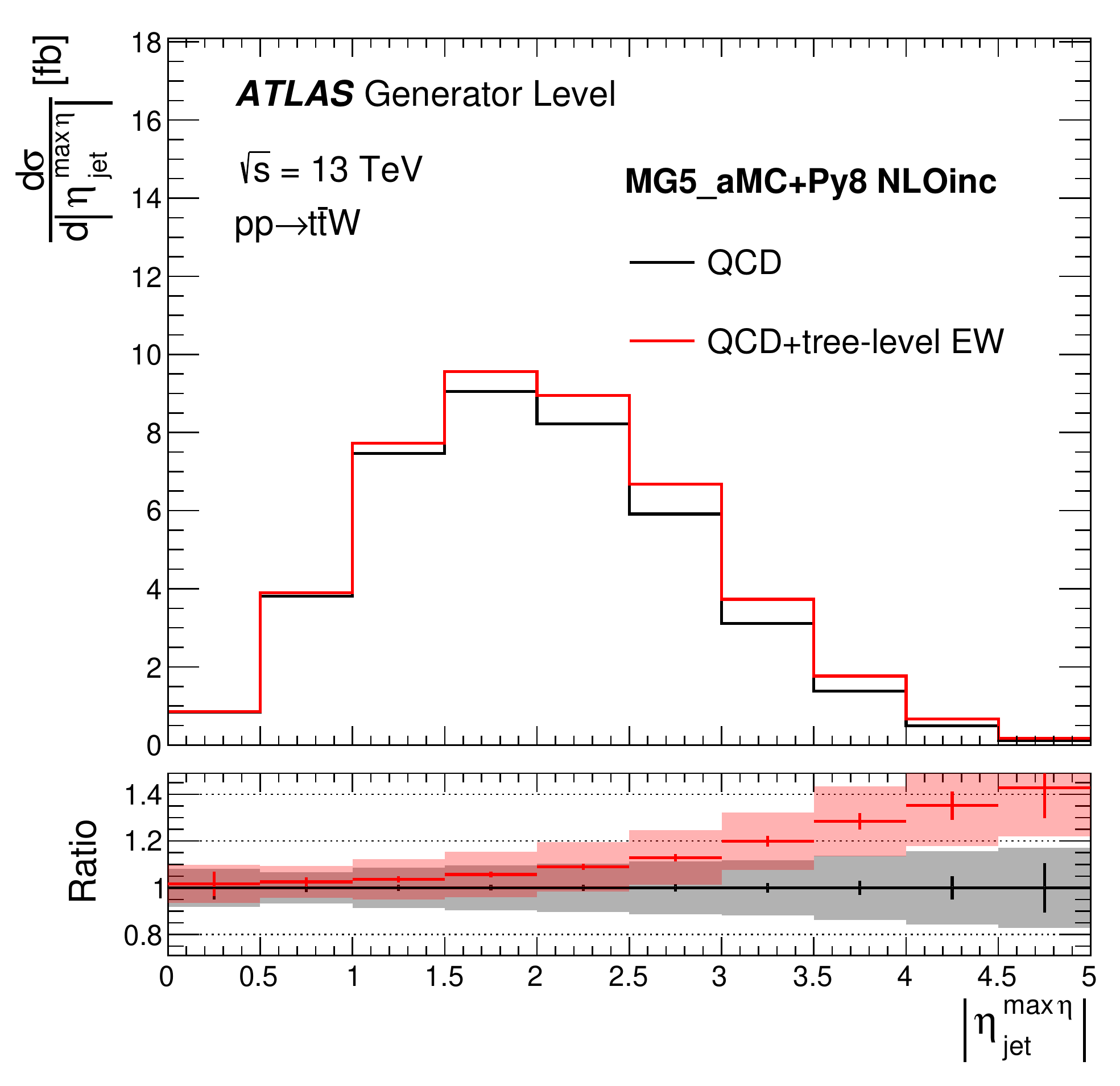} \\
            \footnotesize{(c)} & \footnotesize{(d)} \\
      \end{tabular}
      \caption{(a): Cross--section dependence with two functional forms of the $\mu_R~\mathrm{and}~\mu_F$ scales. (b) to (d): Effect of the ``tree--level EW" contribution for \textsc{MG5\_aMC@NLO} for some kinematic variable distributions. The vertical error lines show the 7--point scale variations for (a), while for the rest they indicate the MC statistical uncertainty and the shaded bands represent these scale variations. Figures from Ref.~\cite{ATLAS:2020esn}.} 
      \label{fig:fig3}
\end{figure}

The EW corrections to the \ttw\ process have been recently calculated to increase the cross--section by around $10\%$~\cite{Frederix:2017wme} which is much bigger than naively expected. This is caused by the appearance of $tW\rightarrow tW$ scattering diagrams $(\mathcal{O}(\alpha_S\alpha^3))$, as those on the bottom right of Figure~\ref{fig:fig1}. Such corrections and their effects are shown in Figure~\ref{fig:fig3} from which the following conclusions may be extracted:
\begin{itemize}
      \item (a): in addition to the strong $\mu_R~\mathrm{and}~\mu_F$ scale dependance, the EW corrections predict a $10\%$ increase in the cross--section throughout for all scale values. A similar study has been performed using \textsc{Sherpa2.2.8} where the effect on the cross--section is of about $5\%$.
      \item (b) to (d): shape effects of around a $20\%$ are observed for events in the high central and forward jet multiplicity regions, as well as in the high pseudo rapidity region ($2.5<|\eta|<4.5$) where the extra jet in $tW\rightarrow tW$ scattering is expected.
\end{itemize}

\section{Conclusions}

From these studies including higher order effects in both QCD and EW it is clear that we still don't have the whole picture for the \ttw\ process. The choice of the functional form of the $\mu_R~\mathrm{and}~\mu_F$ scales as well as their values can change the predictions substantially. The addition of multileg setups and EW corrections also have a $10\%$ impact on the cross--section values. The former seem to be in agreement across different MC generators and also consistent within relevant parameter variations (such as $\mu_Q$); while the latter further increase the cross--section and have considerable shape effects in some kinematic distributions. These results are documented in Ref.~\cite{ATLAS:2020esn}.

\vspace{.5cm}
Copyright 2021 CERN for the benefit of the ATLAS Collaboration. CC-BY-4.0 license.


\end{document}